# Aggregation and visualization of spatial data with application to classification of land use and land cover

Mihal Miu[1], Xiaokun Zhang[1], M. Ali Akber Dewan[1], Junye Wang[1, 2,*]

[1]School of Computing & Information Systems, [2]Faculty of Science and Technology

Athabasca University, 10011, 109 Street, Edmonton, AB T5J 3S8, Canada

Mihal Miu: Tel: +1 6478396994; E-mail address: mihalmiu@hotmail.com

Xiaokun Zhang: Tel: +1 8666039520; E-mail address: xiaokunz@athabascau.ca

Ali Akeber Dewan: Tel: +1 7807612726; E-mail address: adewan@athabascau.ca

* Corresponding author, Junye Wang. Tel: +1 7803944883

E-mail address: junyew@athabascau.ca

## Abstract

Aggregation and visualization of geographical data are an important part of environmental data mining, environmental modelling, and agricultural management. However, it is difficult to aggregate geospatial data of the various formats, such as maps, census and survey data. This paper presents a framework named **PlaniSphere**, which can aggregate the various geospatial datasets, and synthesizes raw data. We developed an algorithm in PlaniSphere to aggregate remote sensing images with census data for classification and visualization of land use and land cover (LULC). The results show that the framework is able to classify geospatial data sets of LULC from multiple formats. National census data sets can be used for calibration of remote sensing LULC classifications. This provides a new approach for the classification of remote sensing data. This approach proposed in this paper should be useful for LULC classification in environmental spatial analysis.

**Key words**: spatial data aggregation, environmental modelling, geospatial mapping, data mapping, land use and land cover classification

## 1. Introduction

With the advancements in remote sensing, monitoring networks and geographic information systems (GIS), the availability of spatial data is rapidly increasing. There currently exist many large repositories of analytical and subject-oriented databases, such as national censuses, statistical frameworks of the UN System of Environmental-Economic Accounting (Yu et al., 2016; UN et al., 2012; Wang et al, 2011; Betrie et al., 2015). These geospatial data include not only maps and locations of land use and land cover (LULC), but also multiple attributes of data, such as socioeconomic data from the census. These data sets are heterogeneous across the various data sources and inconsistent in file formats because different supplier has a tendency to use its proprietary data/file formats. They may have static or dynamic characteristics. Thus, there may be little or no commonality between the formats used. This has led to increased challenges in capturing and analyzing this spatial data.

Various models have been developed for spatially continuous predictions and data analysis in the environmental science and management, such as DeNitrification DeComposition model (DNDC) (Li et al., 2011; 2016; Cardenas et al., 2013; Wang et al., 2012) and the Soil and Water Assessment Tool (SWAT) (Arnold et al., 2012). The accuracy and capacities of these models are dependent on data availability and formats. Therefore, the data transformation has to be carried out for the modelling applications. Most of hydrologic and biogeochemical model data preparations often include significant manual intervention. This is time consuming. Some data preparation tools are developed for some specific models, such as calculation of missing meteorological recordings in SWAT (Arnold et al., 2012). These tools are usually so tightly developed to pre-specified data sources that they may not offer the best appreciated data for a specific region or generally modelling objective. Therefore, a large-scale hydrological or biogeochemical model may not provide proper calibration and uncertainty analyses due to insufficient accurate accounting of vegetation classification, leading to inaccurate model prediction. For example, using long-term soil organic carbon datasets from agricultural field experiments, a process-based model will be limited to drawing on the average of observations or by taking into account the variations in observations in order to predict soil and crop dynamics. Generally speaking, there are three main challenges to large scale environmental modelling data: a) time requirements for transforming data formats due to a lack of automated processes; b) a lack of flexibility in updating available data; and c) difficulties in sharing data with others when creating models (Billah et al., 2016).

Improvements in the use and accessibility of multi-temporal, satellite-derived environmental data or other thematic raster data have contributed to the growing use in environmental modelling (He, 2002). Remote sensing is a data source for hydrological and biogeochemical models because it provides synoptic information on vegetation growth conditions over a large geographic area in near real-time. It is often essential to classify land use and cover for a variety of environmental modelling applications, such as leaf area index, albedo and surface roughness (Delegido et al., 2015; Ju and Masek, 2016; Faramarzi et al., 2015; Baird et al., 2016; Taylora and Lovell, 2012; Khatami et al., 2016; Wang, 2014; Cardenas et al., 2013). The vegetation growth pattern is estimated using the normalized difference vegetation index (NDVI), which is based on visible (red) (VIS) and near-infrared (NIR) band reflectance derived from the most widely used global NDVI data sets (Guay, 2014; Tian 2015). The NDVI expresses the normalized ratio between reflected energy in the red and near infrared regions, and provides an indicator of the vegetation ‘greenness’ (Koltunov et al., 2009; Delegido et al., 2013). However, NDVI values for sparse green vegetation are very similar to those of bare soils or dry vegetation due to the similarity of the spectrum in the VIS and NIR regions for bare soil and dry vegetation (Kokaly et al., 2009), which makes this index unsuitable for distinguishing a green from brown leaf area index (LAI) as well as crop classification. Distortions may be introduced in terms of both cover type quantities and landscape patterns when a classified Landsat Thematic Mapper (TM) satellite image is processed to a vegetation map by using data aggregation. Categorical aggregation generates output values by logical processing such as majority or predominant input categories, or spatial processing such as assigning output values based on the input cell at the centre of the output grid cell. Therefore, its applications will be limited due to quality and cost issues as well as to spatial resolution of land use and cover classifications, particularly for large scale environmental modelling (Boulos, 2005).

Because of the data distortion or even data loss associated with aggregation and scaling and level of heterogeneity across data sources, LULC datasets derived from medium to coarse resolution satellite sensors are particularly inaccurate (Fritz et al., 2010; Fritz and See, 2008). Therefore, it is necessary to have a comprehensive assessment of the remote sensing LULC classification for the performance of predictive models. Efforts have been made to quantify such phenomena and predict its effects across scales. Image data is dependent on its spatial resolution and the ways in which that spatial resolution is handled at the image processing stages, with respect to LULC patches of varying sizes and shapes. The mixture model is common approach in satellite image processing, which uses measurements directly at every scale. Statistical finite mixture models dictate the measurements of some LULC as having come from a finite number of sub-categories of LULC. A LULC probability density can be modelled as the weighted sum of the probability densities of the sub-category crops (Ju and Masek, 2016). Then the multiscale approaches are examined using labels of categories in the field. A patch can be labeled as a general class only when, at a certain spatial scale, it is judged to contain high probability density compared to patches of more than one specific class referring to the level of detail in the categories used in classification. Therefore, field experiments play a key role in obtaining first-hand information about the effects of assessments and calibrations on LULC, crop yield and various carbon and nitrogen pools or fluxes in particular fields. However, most field sampling and labeling experiments require large amounts of time and resources. As a consequence, field observations will always be restrained in space and time for practical and financial reasons. Although LULC classifications can be estimated with a certain degree of reliability using the limited samples of ground field observations, details on exact LULC and the spatial distribution are rarely available for calibrations of the processing algorithm of remote sensing images (Fritz et al., 2010). The lack of thorough filed sampling or labeled field data poses an obstacle at regional scales. Therefore, it has been a challenge to validate the remote sensing modelling of LULC classification in large-scale region.

Generally speaking, a big-data, including LULC, has been created through various national censuses in the world. These rich data stimulated our idea to explore a new approach for the calibration and validation of remote sensing LULC classification using national census data. To the best of our knowledge, no such approaches have been reported. Although some software of GIS, such as ArcGIS, is able to aggregate spatial data, LULC classification of remote sensing images is complicated as discussed above. Users usually need to develop special algorithms and models to calibrate remote sensing images against field observation data. There are no such functions in existing GIS software. Furthermore, ArcGIS is not an open source platform.

The objective of this paper is to develop an algorithm in PlaniSphere framework (PlaniSphere, 2016) for aggregation and visualization of LULC classification. National census data is used for validation and calibration of classification of LULC of remote sensing. Thus, classification of LULC of remote sensing image can be performed in regional scale since many censuses are available at the national scale. This will improve the accuracy of LULC classifications using conventional approaches of limited field labeling and sampling at small scale.

## 2. Methodology

### 2.1. PlaniSphere framework

Many analyzing techniques and the increasing availability of geo-referenced data could provide an effective way to manage spatial information by providing large-scale storage, and multidimensional data management together in one system (Herbreteau et al., 2007; Boulil et al., 2015; Sarwat, 2015).

PlaniSphere developed by Miu (2016) is a desktop platform with graphical capabilities to aggregate spatial data with different formats. The software tool can aggregate files from local storage that will conform to widely used file formats such as: GeoTIFF files, ESRI shape files and KML files (Figure 1). A graphical user interface (GUI) has been developed with the Plug-In Manager, as shown in Figure 2. This class is responsible for creating instances and loading jar files that support the functionality of the plug-in.

See Figure 2 for the general layout of the main window. Each file will represent a single layer that will be rendered on the map. A ribbon band will be dedicated to “Custom Map Sources” support. It should be noted that each time a connection to a known map server is required the user will need to click on the “Add WMS Server” button in the ribbon band. Each file type will be represented by a button. A high-level transform from local storage or from remote servers can be implemented through the Internet using Web Map Service protocol (WMS, 2016). It is an application that revolves around a main window with a ribbon toolbar. The toolbar will present the core features to the user. Under the ribbon there will be a working area where a map or 3D virtual globe image can be displayed. The optional capability to manipulate any layers within the map will be in a tabbed pane to the left of the working area.

### ➢ 2.2. LULC algorithm

LULC classification and mapping of satellite or airborne images have increased exponentially in the recent decades because of improved data availability and accessibility (Yu et al., 2014; Ju and Masek, 2016). LULC classification and mapping are complicated processes that converts remotely sensed imagery into usable data. Considerable efforts have been undertaken to improve LULC classification accuracy, such as with the spatial and temporal distribution of terrestrial net primary production (Parazoo et al., 2014; Cleveland et al., 2015), and leaf area index (Chen and Cihlar, 1996). For example, an urban planning requires to distinguish the LULC among the infrastructure, such as roads, buildings, houses, rail and so on. Maimaitijiang et al. (2015) studied spatial and temporal dynamics of urban growth in the St. Louis metropolitan region using remote sensing derived LULC changes and socio-economic factors. This requires a higher level of land analysis than simple analysis of urbanization (Jacobs-Crisioni et al., 2014). Taylor and Lovell (2012) studied urban agriculture to category the level of urbanization and urban agriculture. They observe green colours from non-green colours. Green represented vegetation while the non-green colours represent the developed areas. However, due to limited reference sites, the accuracy is commonly limited to single sites on reference data from field sampling. Such references are too limited to infer general guidelines for selecting a suitable process to produce highly accurate maps (Stehman, 2006).

In one of our experiments, Plug-In manager of PlaniSphere was created for user programming. The current approach is to use census data as a labeling or sampling field for calibration and validation of conventional classification of pixel-wise remote sensing images. The solution proposed is that colour matching be performed. It is assumed that each vegetation type will have its distinct colour. A carefully analysis of distinct areas represented by different colours reveals that each area does not have a single distinct colour. The correct statement would be to say that an area has a distinct colour range. This will allow for a colour range to be mapped to a particular vegetation type/subtype. Once the colour range matching is performed, areas occupied by each vegetation type can be calculated. This can be compared with existing census data. If the resulting numbers are a match than the colour range matching works. If not than the colour ranges used will need to be redefined. All of this is demonstrated in the land analysis plug-in that has been created for PlaniSphere. An algorithm in PlaniSphere Plug-In was developed for the LULC classification of remote sensing images with aggregation of survey or census data. Figure 3 illustrates the flow chart of the algorithm for LULC classification of remote sensing. A scheme can be written based on the colour spectrum using an xml file for each type of crops and code procedure using JAVA language in Plug-In land use manager as follows:

```
<RangeList>
    <Range>
            <Name>Band 1 - Wheat</Name>
<Comment>Band 1</Comment>
<Color>606f55</Color>
<Tolerance>10</Tolerance>
    <Range>
    <Range>
            <Name>Band 2 - Canola</Name>
<Comment>Band 2</Comment>
<Color>897966</Color>
<Tolerance>10</Tolerance>
```

<Range>

<Range>

<Name>Band 3 – Misc Crop</Name>

<Comment>Band 3</Comment>

<Color>a59385</Color>

<Tolerance>10</Tolerance>

<Range>

<Range>

<Name>Band 4 - Pasture</Name>

<Comment>Band 4</Comment>

<Color>5f6655</Color>

<Tolerance>10</Tolerance>

<Range>

<Range>

<Name>Band 5 - Canola</Name>

<Comment>Band 5</Comment>

<Color>515546</Color>

<Tolerance>10</Tolerance>

<Range>

<Range>

<Name>Band 6 - Wheat</Name>

<Comment>Band 6</Comment>

<Color>918070</Color>

<Tolerance>10</Tolerance>

<Range>

```
<Range>
        <Name>Band 7 - Pulses</Name>
<Comment>Band 7</Comment>
<Color>988775</Color>
<Tolerance>10</Tolerance>
    <Range>
    <Range>
```

The above program represents an association between colour bands and area of each crop type. The program calculates area of every crop type automatically to compare with its census data and analyse errors between the calculated area and census data. When the calculated area of each crop types matches its census data within its tolerance, the program stops.

### □ 2.3. LULC Implementation

The input data of LULC are satellite remote sensing images and census data (Table 1). The images can be GeoTIFF files, ESRI shape files and KML files (Figure 1). Figure 4 is a typical of a Bing image map with a latitude and longitude coordinate which is used to test the algorithm. Overlaid on the map is the content from an ESRI shapefile. The shape file contains parish/township boundaries delimited by red lines. The boundary information will aid in selecting an area used for land analysis of classifications. Since statistical data of LULC exists in each of the parish/townships they can be used for the classifications of LULC in the image. A graticule is used as the top most layer in order to display parallels and meridians used for navigation purposes. The two layers required for analysis are the Bing image layer and the shapefile layer containing the parish/township boundaries. The graticule is optional, it is not necessary for analysis but it helps by identifying areas (based on latitude and longitude) and navigation from one area to another.

Colour analysis requires the GeoTIFF export function provided by PlaniSphere and the plug-in infrastructure. The GeoTIFF export function generates a 32-bit colour depth GeoTIFF, where 8-bits are used to represent red colours, 8-bits are used to represent green colours, 8-bits are used to represent blue colours and 8-bits used for the alpha channel used for transparency. Each colour range needs to have a used defined name, a range of values (this is the colour range) and an optional user defined comment. It is proposed that all of this be represented by an XML file. It should be noted that XML files can be easily created using a simple text editor in Plug-In manager. The scheme represents detailing the colour map to be used by the land analysis plug-in.

LULC classification will be performed using a three-step process (Figure 3). The first step is to aggregate geospatial data. This step will produce a map (graphical image) where a pixel analysis can be performed. The second step is to aggregate census data based on the same geographic boundaries used to aggregate geospatial data in the first step. The third step is to compare the data generated from the first step with the data from the second step. The first two steps demonstrate the capacity of data aggregation. The third step demonstrates the ability to compare data and, with user intervention, the ability to improve or complement the data.

## 3. Results and Discussions

Analysis of LULC is performed on two parishes/townships (named TO39R20W4 and TO38R21W4) in the rural Alberta (Figure 4). The parish/township, TO39R20W4, is located in North West of the Town of Stettler (Red colour) (Figure 5), south of Highway 601 and east of Highway 835. The parish/township named TO38R21W4 (Yellow colour) can be found south of highway 12, north of Township Rd 382, East of Range Rd 220 and west of Range Rd 210 (Figure 5) .

A colour map has been created (Figure 6) and it will be used for analysis of the parish/township TO39R20W4. The GeoTiff image dimensions are at 2045 (width) × 2048 (height) for TO38R21W4. GeoTiff resolution is that one pixel represents $2.23^{-5}$ $km^2$, and total GeoTiff area is 93 $km^2$.

Figure 7 show a map generated by Land Analysis Plug-in by implementing the above algorithm. Table 1 provides statistical data stating mean (average) values for the 2009 to 2012 periods. The pixel analysis technique was able to detect bodies of water (lakes, reservoirs, etc.). However, the census data does not describe bodies of water. Accordingly, this result incorporating the pixel analysis technique, can be used to enhance findings or generate new census data based on the remote analysis. It should be understood that the information provided in Table 1 is an illustration and may not be 100% accurate. The association between the crop types and colour bands represents an educated guess using a comparison of the statistical data (Table 1) (NRC, 2016) and the fact that each colour may represent a type of plant or a type of field. For example, *wheat*: a wheat field when it is planted may be brown since the seeds are inserted into the bare ground. As the wheat starts to grow, it will appear as a green field from an aerial photograph. When the wheat matures it changes from green to a golden or brown colour. Hence, planted fields can not only be represented by green shades but also by brown and yellow shades visible in the aerial images. This is another argument for using colour infrared imagery to accurately determine crop types due to seasonal change. This is to some extent arbitrarily selecting filtering and smoothing algorithms during representing crop index (Chen et al., 2014). However, these issues can be improved using weighted colour for each type crop in modelling, such as mixture model (Ju, et al., 2005).

The use of satellite imagery is common in large scale environmental modelling. Since national censuses are performed regularly, these databases provide a potential solution for LULC classification and labeled calibration requirements on a regional scale. In particular, the Intergovernmental Panel on Climate Change (IPCC, 2016) recommends compiling IPCC inventory of greenhouse gas emission based on a database of the national census while many models on climate change use remote sensing imaging (Osborne et al., 2015; Arnold et al., 2012). The input data may be inconsistent as a result of the different data sources. Because of the considerable work required for reference data creation and the limited scope of most studies, accuracy of results reported in these studies is usually limited to single sites with testing performed on reference data from a single image. Such comparisons are too limited to infer general guidelines for selecting suitable processes to produce highly accurate maps (Stehman, 2006).

The proposed method may have the potential to unify these two approaches. The results demonstrate the capacity of the recommended approach by comparing the accuracy of the proposed classification processes with that of the existing census data. Comparing the results from the land analysis plug-in to the data in Table 1 illustrates the difficulty of correlating this information. A comparison would not be appropriate as Table 1 provides statistical mean values for the years between 2009 and 2012 (NRC, 2016).

However, like an NDVI analysis, the values for sparse green vegetation are very similar to those of bare soils or dry vegetation due to the similarity of the colour spectrum for bare soil and dry vegetation (Kokaly et al., 2009). This limits the accuracy of distinguishing green from brown LAI while vegetation is being classified. Distortions may be introduced by data aggregation in terms of both cover type quantities and landscape patterns when using a vegetation map based on a classified satellite image. Unlike conventional NDVI, the present approach uses statistical data to examine and calibrate the algorithm. The effects of this distortion can be improved. Because national level censuses, including crop production, are carried out regularly at national and regional scales (Statistics Canada, 2016), these rich census data are potential as a new approach for calibration and validation of remote sensing classification.

The method proposed in this paper is an automated process to transforming data formats since an image format is an option. Thus, once updating data (images and census data) is available, the algorithm is able to implement the data aggregation and sharing data is possible.

However, the present land analysis plug-in analyzes an image for a very specific time, which is a certain day of a certain year, while the exact date of the Bing images may be unknown. A fair comparison would be to obtain statistical data based on median values for a period. Aerial images should be obtained for the same period. Analysis would be performed on all of the images and a median comparison could be performed between the values obtained from analyzing the aerial images and the statistical data. It should be noted that using the mean for analysis has a major disadvantage. Mean analysis is susceptible to outliers (values that are unusual compared to the rest of the data set). These problems stem from using a collection of windows that induce a fixed partitioning of the image and data sources, but they can be overcome with the national censuses provided for the specific year (Statistics Canada, 2016), and images provided by Google Earth for the specific time over many regions. Furthermore, advanced technologies such as Lidar, which can greatly improve resolution, have become more economical and more generally available for these applications. A model, such as mixture model (Ju et al., 2005; Ju and Masek, 2016), would allow for the use of an enlarged collection of windows corresponding to a fully redundant system of all possible partitioning for data collection and averaging of the representations from these partitionings. The weight factors of the mixture model can be replaced by the percentage of the census data in the present method. Our future work will develop a mixture model to improve accuracy of the current method and a larger region will be tested.

## 4. Conclusions

Despite the emergence of significant remote sensing and modelling capabilities, field observations are often restrained in space and time for practical and financial reasons. Therefore, it has been a challenge to validate the remote sensing modelling of LULC classification in large-scale region. Much remains to be done before the full knowledge of multiscale phenomena occurring in LULC can be directly utilized in LULC classification of remote sensing images and the development of environmental modelling. In this paper, we develop a new algorithm in PlaniSphere framework to calibrate and validate remote sensing data with census data as the first attempt. The plugin function of PlaniSphere is used for classification of LULC, by providing customizations to serve uses. The results show that this algorithm and framework can aggregate spatial data and LULC census data (available in quantity in the national database) to create a LULC classification map that includes geospatial data sets from multiple sources. We demonstrate potential applications of census data for the calibration and validation of LULC classification of the remote sensing images. LULC from remote sensing images can be classified and calibrated using existing census data. This enhances environmental informatics and data analytics for large scale modelling of hydrological and biogeochemical processes. Because of the rich data resource of census, this can offer a new approach for validation and calibration of remote sensing land classification. The methods proposed in this paper can also be useful for a variety of scale-related LULC classification tasks for other GIS platforms, such as ArcGIS.

**Acknowledgements**

M.M. would like to thank James R. Miller, Department of Electrical Engineering and Computer Science, The University of Kansas for providing access to Lidar Data Visual Analysis and Editing. http://people.eecs.ku.edu/~miller/

M.M. would like to thank the City of Vancouver for LiDAR data collected in 2013 with coverage up to extents of City of Vancouver's legal jurisdiction. http://data.vancouver.ca/datacatalogue/LiDAR2013.htm

## References

Arnold, J.G., Moriasi, D.N., Gassman, P.W., Abbaspour, K.C., White, M.J., Srinivasan, R., Santhi, C., Harmel, R.D., van Griensven, A., Van Liew, M.W., Kannan, N., Jha, M.K.. 2012. SWAT: Model Use, Calibration, and Validation. Transactions of the ASABE. 55(4), 1491-1508.

Baird, M.E., Cherukuru, N., Jones, E., Margvelashvili, N., Mongin, M., Oubelkheir, K., Ralph, P.J., Rizwi, F., Robson, B.J., Schroeder, T., Skerratt, J., Steven, A.D.L., Wild-Allen, K.A. Remote-sensing reflectance and true colour produced by a coupled hydrodynamic, optical, sediment, biogeochemical model of the Great Barrier Reef, Australia: Comparison with satellite data. Environmental Modelling & Software 78 (2016) 79-96.

Betrie, G.D., Deng B., Wang, J., 2015. Integrated Modeling of the Athabasca River Basin using SWAT, in: Chang, M., Al-Shamali, F. (Eds.), 2015 Proceedings of Science and Technology Innovations, Chapter 3, Athabasca University, Athabasca, pp. 27-38.

Bhatt, G., Kumar, M., Duffy, C. J., 2014. A tightly coupled GIS and distributed hydrologic modeling framework. Environ. Model. Softw. 62, 70–84.

Billah, M.M., Goodall J.L., Narayan, U., Essawy, B.T., Lakshmi, V., Rajasekar, A., Moore, R.W., 2016. Using a data grid to automate data preparation pipelines required for regional-scale hydrologic modeling, Environmental Modelling & Software 78, 31-39.

Boulil, K., Bimonte, S., Pinet, F., 2015. Conceptual model for spatial data cubes: A UML profile and its automatic implementation, Computer Standards & Interfaces 38, 113-132.

Boulos, M.N., 2005. Web GIS in practice III: creating a simple interactive map of England's Strategic Health Authorities using Google Maps API, Google Earth KML, and MSN Virtual Earth Map Control. Int. J. Health Geogr. 4, 22. DOI: 10.1186/1476-072X-4-22

Cardenas, L.M., Gooday, R., Brown, L., Scholefield, D., Cuttle, S., Gilhespy, S., Matthews, R., Misselbrook, T., Wang, J.Y., Li, C., Hughes, G., Lord, E., 2013. Towards an improved inventory of $N_2O$ from agriculture: Model evaluation of N2O emission factors and N fraction leached from different sources in UK agriculture, Atmospheric Environment 79, 340-348.

Chen, J.M., Cihlar, J., 1996. Retrieving Leaf Area Index of Boreal Conifer Forests Using Landsat TM Images. Remote Sensing of Environment 55(2), 153-162.

Chen, W., Foy, N., Olthof, I., Zhang, Y., Fraser, R., Latifovic, R., Poitevin, J., Zorn, P., McLennan, D., 2014. A biophysically based and objective satellite seasonality observation method for applications over the Arctic. International Journal of Remote Sensing 35(18), 6742–6763.

Cleveland, C.C., Taylor, P., Chadwick, K.D., Dahlin, K., Doughty, C.E., Malhi, Y., Smith, W. K., Sullivan, B.W., Wieder, W.R., Townsend, A.R., 2015. A comparison of plot-based satellite and Earth system model estimates of tropical forest net primary production, Global Biogeochemical Cycles 29(5), 626-644.

Delegido, J., Verrelst, J., Meza, C.M., Rivera, J.P., Alonso, L., Moreno, J., 2013. A red edge spectral index for remote sensing estimation of green LAI over agroecosystems. Eur. J. Agron. 46, 42–52.

Delegido, J., Verrelst, J., Rivera, J.P., Ruiz-Verdú, A. Laboratorio, J.M., 2015. Brown and green LAI mapping through spectral indices, International Journal of Applied Earth Observation and Geoinformation 35, 350–358.

ESRI team, 2016, ArcGIS software, available online https://www.arcgis.com/features/ [accessed 28.06.16]

Faramarzi, M., Srinivasan, R., Iravani, M., Bladon, K.D., Abbaspour, K.C., Zehnder, A.J.B., Goss, G.G., 2015. Setting up a hydrological model of Alberta: Data discrimination analyses prior to calibration, Environmental Modelling & Software 74, 48-65.

Fritz, S., See, L., 2008. Identifying and quantifying uncertainty and spatial disagreement in the comparison of global land cover for different applications. Glob. Change Biol. 14 (5), 1057-1075.

Fritz, S., See, L., Rembold, F., 2010. Comparison of global and regional land cover maps with statistical information for the agricultural domain in Africa. Int. J. Remote Sens. 31 (9), 2237-2256.

Guay, K.C., Beck, P.S.A., Berner, L.T., Goetz, S.J., Baccini, A., Buermann, W., 2014. Vegetation productivity patterns at high northern latitudes: a multi-sensor satellite data assessment, Global Change Biology 20, 3147–3158.

Hallgren, W., Beaumont, L., Bowness, A., Chambers, L., Graham, E., Holewa, H., Laffan, S., Mackey, B., Nix, H., Price, J., Vanderwal, J., Warren, R., Weis, G., 2016. The Biodiversity and Climate Change Virtual Laboratory: Where ecology meets big data, Environmental Modelling & Software 76, 182-186.

He, H.S., Ventura, S.J., Mladennff, D.J., 2002. Effects of spatial aggregation approaches on classified satellite imagery, Int. J. Geographical Information Science 16(1), 93-109.

Herbreteau, V., Salem, G., Souris, M., Hugot, J.P., Gonzalez, J.P., 2007. Thirty years of use and improvement of remote sensing, applied to epidemiology: from early promises to lasting frustration. Health & Place, 13(2), 400-403.

Horsburgh, J.S., Tarboton, D.G., Piasecki, M., Maidment, D.R., Zaslavsky, I., Valentine, D., Whitenack, T., 2009. An integrated system for publishing environmental observations data. Environ. Modell. Softw. 28 (4), 879-888.

IPCC, 2016. Available online http://www.ipcc.ch/ [accessed 28.06.16]

Jacobs-Crisioni, C., Rietveld, P. Koomen, E., 2014. The Impact of Spatial Aggregation on Urban Development Analysis, Applied Geography 47, 46–56.

Ju, J., Gopala, S., Kolaczyk E.D., 2005. On the choice of spatial and categorical scale in remote sensing land cover classification. Remote Sensing of Environment 96, 62–77.

Ju, J., Masek, J.G., 2016. The vegetation greenness trend in Canada and US Alaska from 1984–2012 Landsat data. Remote Sensing of Environment 176, 1–16.

Kokaly, R.F., Asner, G.P., Ollinger, S.V., Martin, M.E., Wessman, C.A., 2009. Char-acterizing canopy biochemistry from imaging spectrometer data for study in gecosystem processes. Remote Sens. Environ. 113, S78–S91.

Khatami, R., Mountrakis, G., Stehman, S.V., 2016. A meta-analysis of remote sensing research on supervised pixel-based land-cover image classification processes: General guidelines for practitioners and future research. Remote Sensing of Environment 177, 89–100.

Kamadjeu, R., 2009. Tracking the polio virus down the Congo River: a case study on the use of Google Earth in public health planning and mapping. Int. J. Health Geogr. 8, 4.

Koltunov, A., Ustin, S.L., Asner, G.P., Fung, I., 2009. Selective logging changes forest phenology in the Brazilian Amazon: evidence from MODIS image time series analysis. Remote Sens. Environ. 113, 2431–2440.

Li, J., Heap, A.D., Potter, A., Daniell, J., 2011. Application of machine learning methods to spatial interpolation of environmental variables. Environ. Model. Softw. 26, 1647-1659.

Lidar, 2013, City of Vancouver, http://data.vancouver.ca/datacatalogue/LiDAR2013.htm [accessed 28.06.16]

Maimaitijiang, M., Ghulam, A., Onésimo Sandoval, J.S., Maimaitiyiming, M., 2015. Drivers of and land cover changes in St. Louis metropolitanarea over the past 40 years characterized by remote sensing andcensus population data, International Journal of Applied Earth Observation and Geoinformation 35, 161–17.

Miller, M., Odobasic, D., Medak, D., 2010. An Efficient Web-GIS Solution based on Open Source Technologies: A Case-Study of Urban Planning and Management of the City of Zagreb, Croatia, FIG Congress 2010, Facing the Challenges – Building the Capacity, April 2010, http://www.fig.net/pub/fig2010/papers/ts05b%5Cts05b_miler_odobasic_et_al_4291.pdf [accessed 28.06.16]

Miu, M., 2016. Aggregation of map (geospatial) data, Athabasca University, Master thesis.

NRC (Natural Resources Canada), 2016. Available online http://www.nrcan.gc.ca/earth-sciences/geography/topographic-information/free-data-geogratis/faq/17284 [accessed 28.06.16]

Osborne, T., Gornall, J., Hooker, J., Williams, K., Wiltshire, A., Betts, R., and Wheeler, T., 2015, JULES-crop: a parametrisation of crops in the Joint UK Land Environment Simulator, Geosci. Model Dev. 8, 1139-1155

Parazoo, N.C., Bowman, K., Fisher, J.B., Frankenberg, C., Jones, D.B.A., Cescatti, A., Pérez-Priego, O., Wohlfahrt, G., Montagnani, L., 2014. Terrestrial gross primary production inferred from satellite fluorescence and vegetation models, Global Change Biology 20, 3103–3121.

PlaniSphere, 2016, www.planisphere.biz [accessed 28.06.16]

Sarwat, M., 2015. Interactive and Scalable Exploration of Big Spatial Data -- A Data Management Perspective, Mobile Data Management (MDM), 2015 16th IEEE International Conference, Volume 1, pp. 263-270, Pittsburgh

Shu, Y., Liu, Q., Taylor, K., 2016. Semantic validation of environmental observations data, Environmental Modelling & Software 79, 10-21.

SRTM (Shuttle Radar Topography Mission), 2016. The Mission to Map the World, Jet Propulsion Laboratory, California Institute of Technology, http://www2.jpl.nasa.gov/srtm/ [accessed 28.06.16]

Statistics Canada, 2016. Available online http://www5.statcan.gc.ca/subject-sujet/theme-theme.action?pid=920&lang=eng&more=0 [accessed 28.06.16]

Taylor, J.R., Lovell, S.T., 2012. Mapping public and private spaces of urban agriculture in Chicago through the analysis of high-resolution aerial images in Google Earth, Landscape and Urban Planning 108, 57– 70.

Tian, F., Fensholt, R., Verbesselt, J., Grogan, K., Horion, S., Wang, Y., 2015. Evaluating temporal consistency of long-term global NDVI datasets for trend analysis, Remote Sensing of Environment, 163, 326–340.

Stehman, S.V., 2006. Design, analysis, and inference for studies comparing thematic accuracy of classified remotely sensed data: A special case of map comparison. Journal of Geographical Systems, 8(2), 209–226.

United Nations, European Commission, Food and Agriculture Organization of the United Nations, International Monetary Fund, Organisation for Economic Cooperation and Development, World Bank, 2012. System of Environmental-Economic Accounting 2012: Central Framework, prepublication (white cover). http://unstats.un.org/unsd/envaccounting/White_cover.pdf. [accessed 28.06.16]

Wang, J., 2014. Decentralized biogas production and farm ecosystem: opportunities and challenges, Frontier in Energy Research 2 (10), 1-12. http://dx.doi.org/10.3389/fenrg.2014.00010

Wang, J., Cardenas, L.M., Misselbrook, T.H., Gilhespy, S., 2011. Development and application of a detailed inventory framework of nitrous oxide and methane emissions from agriculture, Atmospheric Environment, 45(7), 1454-1463.

Wang, J., Cardenas, L.M., Misselbrook, T.H., Li, C.S., 2012. Modelling nitrous oxide emissions in grazed grassland systems. Environmental Pollution. 162, 223-233.

WMS (Web Map Service), 2016. Open Geospatial Consortium, available online. http://www.opengeospatial.org/standards/wms

Yu, L., Liang, L., Wang, J., Zhao, Y., Cheng, Q., Hu, L., et al., 2014. Meta-discoveries from a synthesis of satellite-based land-cover mapping research. International Journal of Remote Sensing, 35(13), 4573–4588.

Figure Captions

Figure 1. Various file formats and protocols used by geospatial data vendors/suppliers.

Figure 2. PlaniSphere main GUI window with the Plug-in Manager and Sample Plug-ins.

Figure 3. Parishes/Townships used in land analysis

Figure 4. Overview diagram of land cover analysis and flow chart

Figure 5. Two Parishes/Townships location

Figure 6. A LandSat photograph of land use and land cover for Parish TO39R20W4

Figure 7. Maps generated by Land Analysis Plug-in by implementing Algorithm

Table Captions

Table 1: A comparison between pixel analysis data and statistical data.

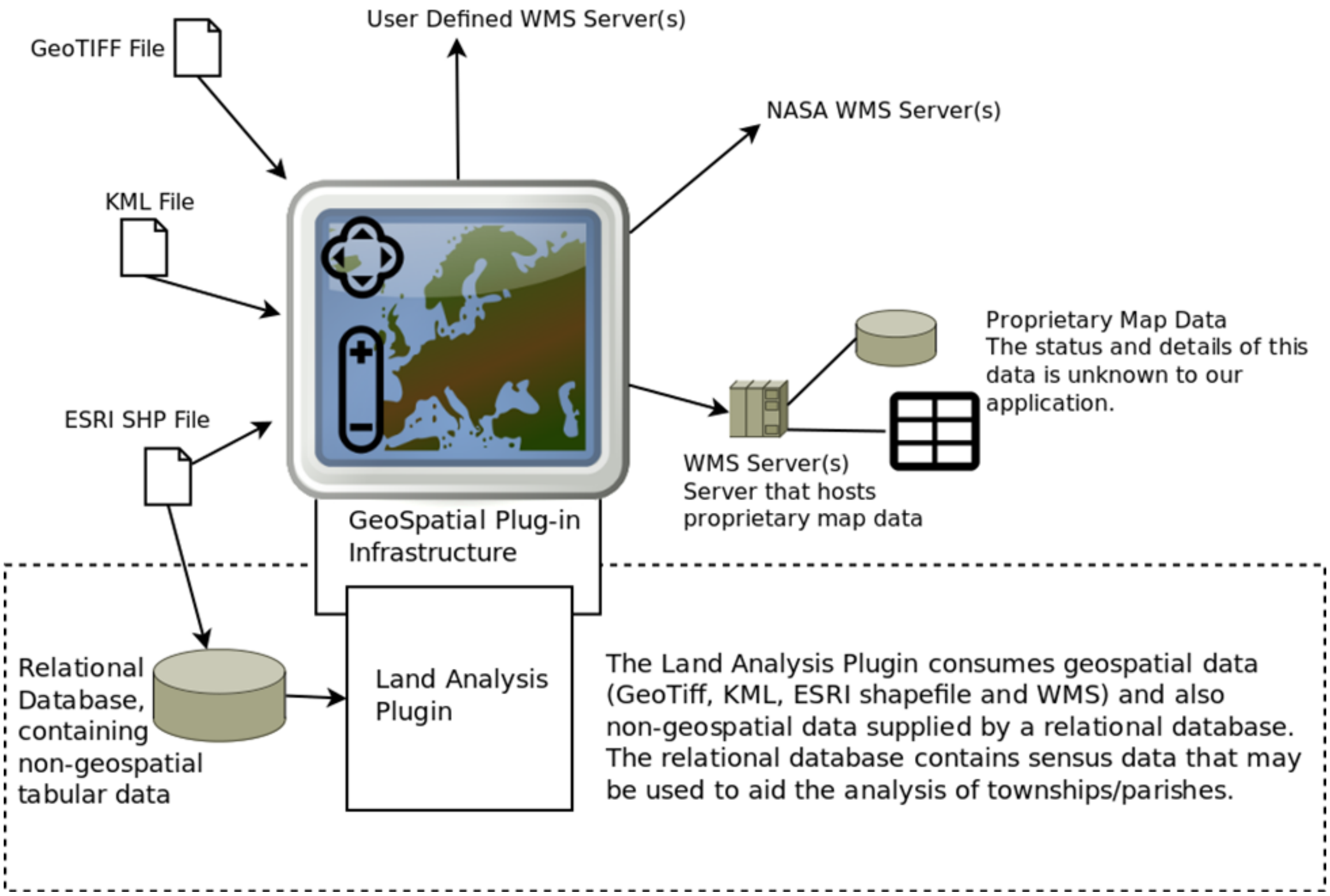

Figure 1. Various file formats and protocols used by geospatial data vendors/suppliers.

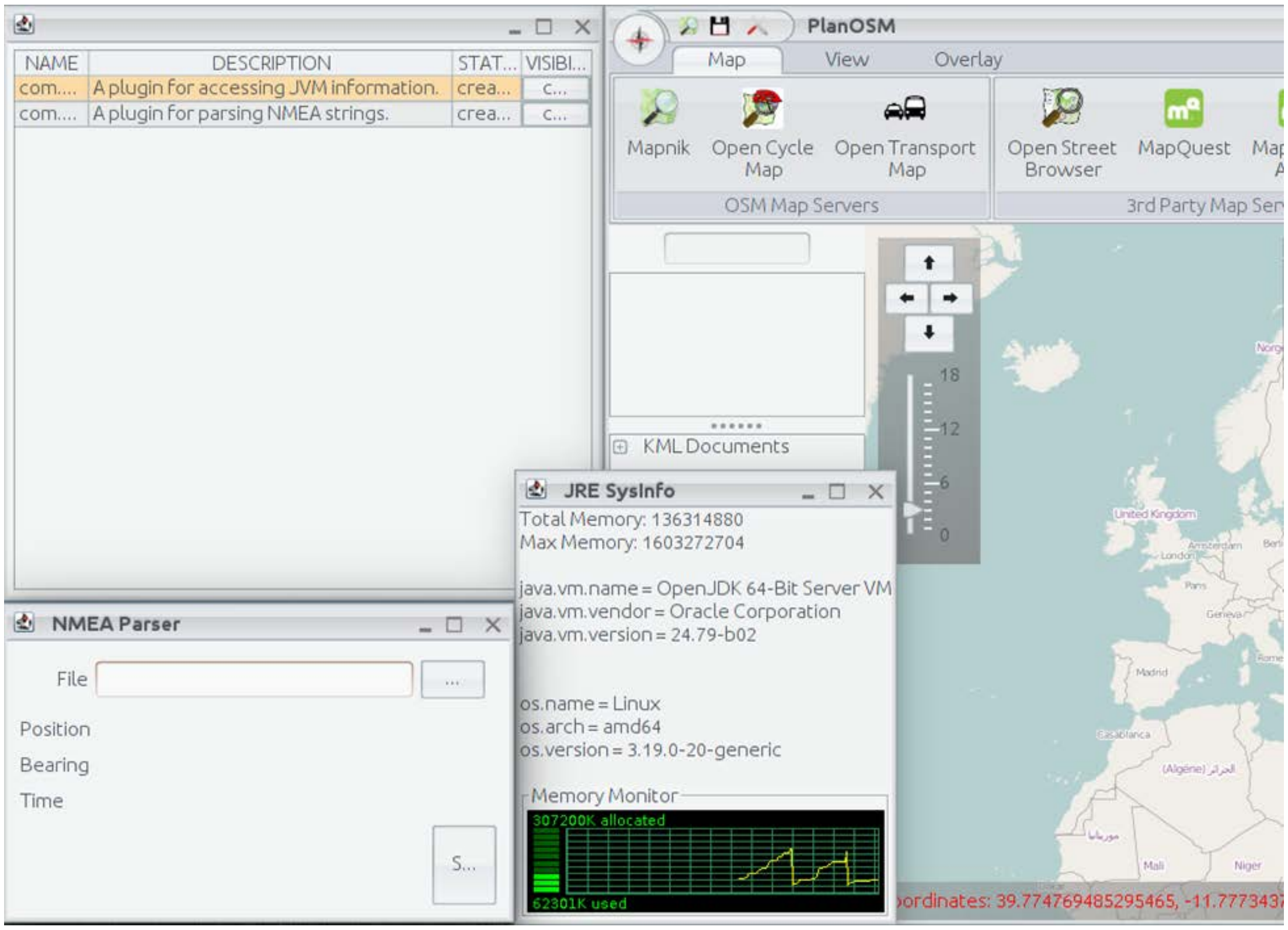

Figure 2. PlaniSphere main GUI window with the Plug-in Manager and Sample Plug-ins.

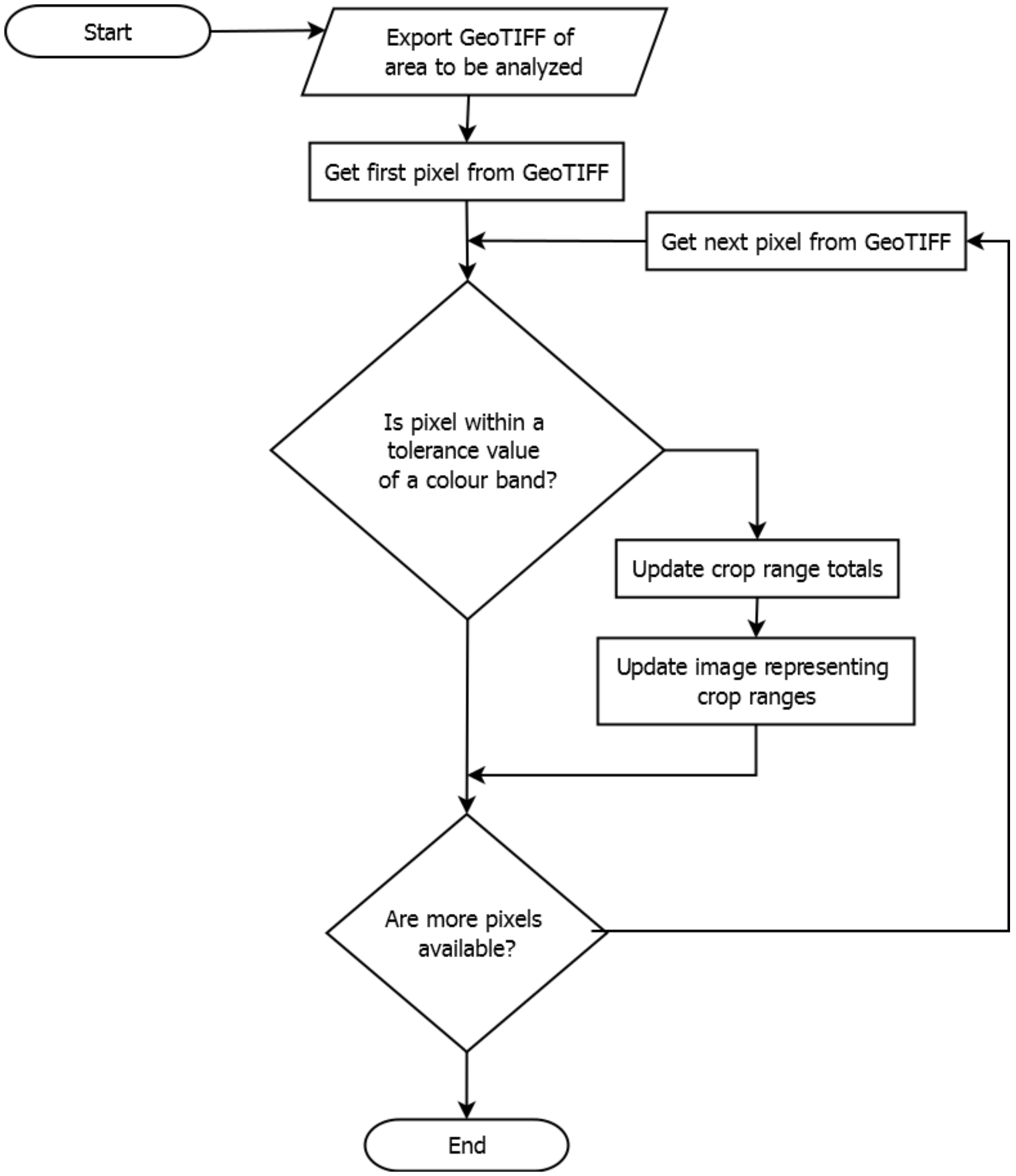

Figure 3. Overview diagram of land use and cover analysis and flow chart

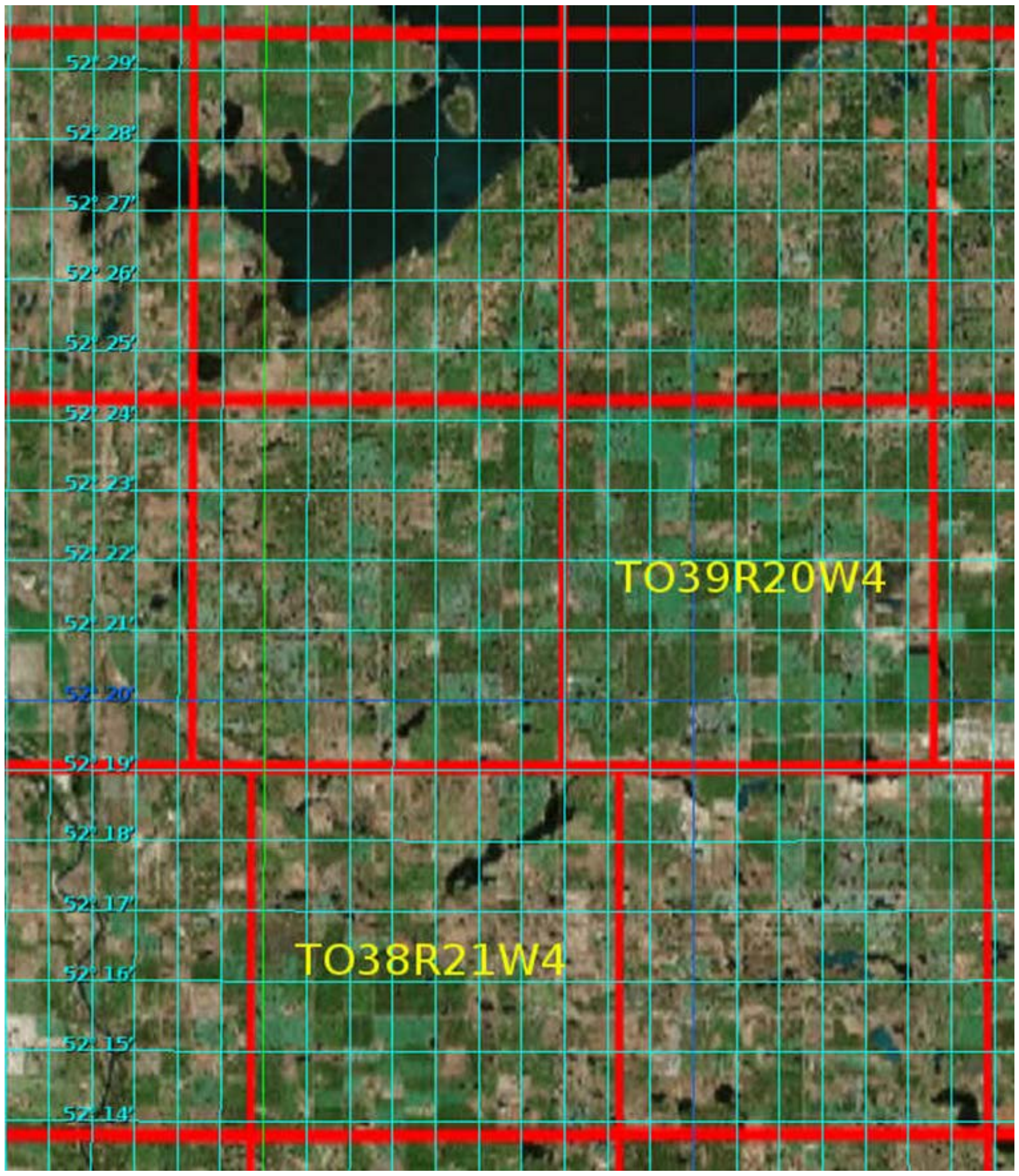

Figure 4. Parishes/Townships used in land analysis

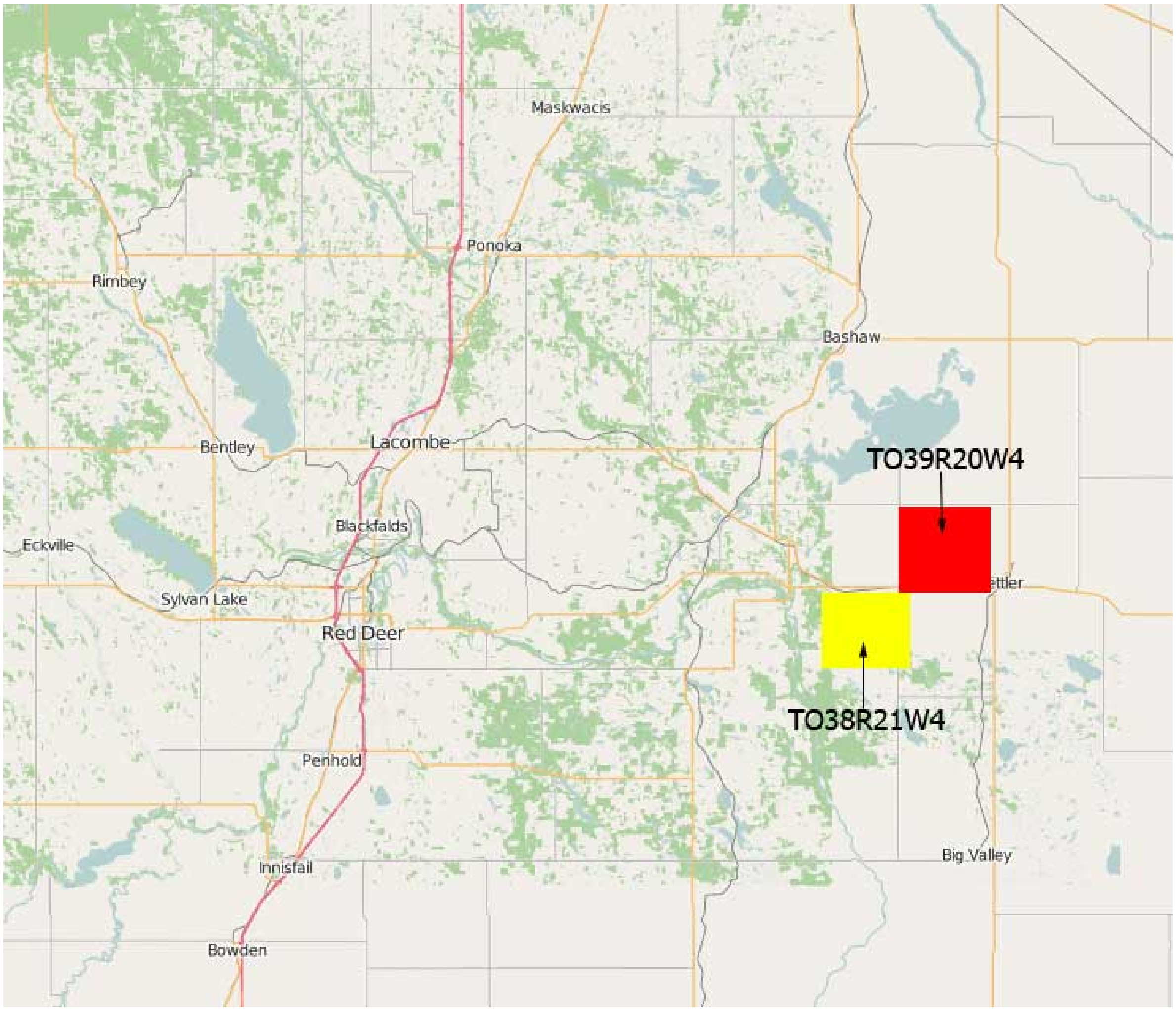

Figure 5. Two Parishes/Townships location

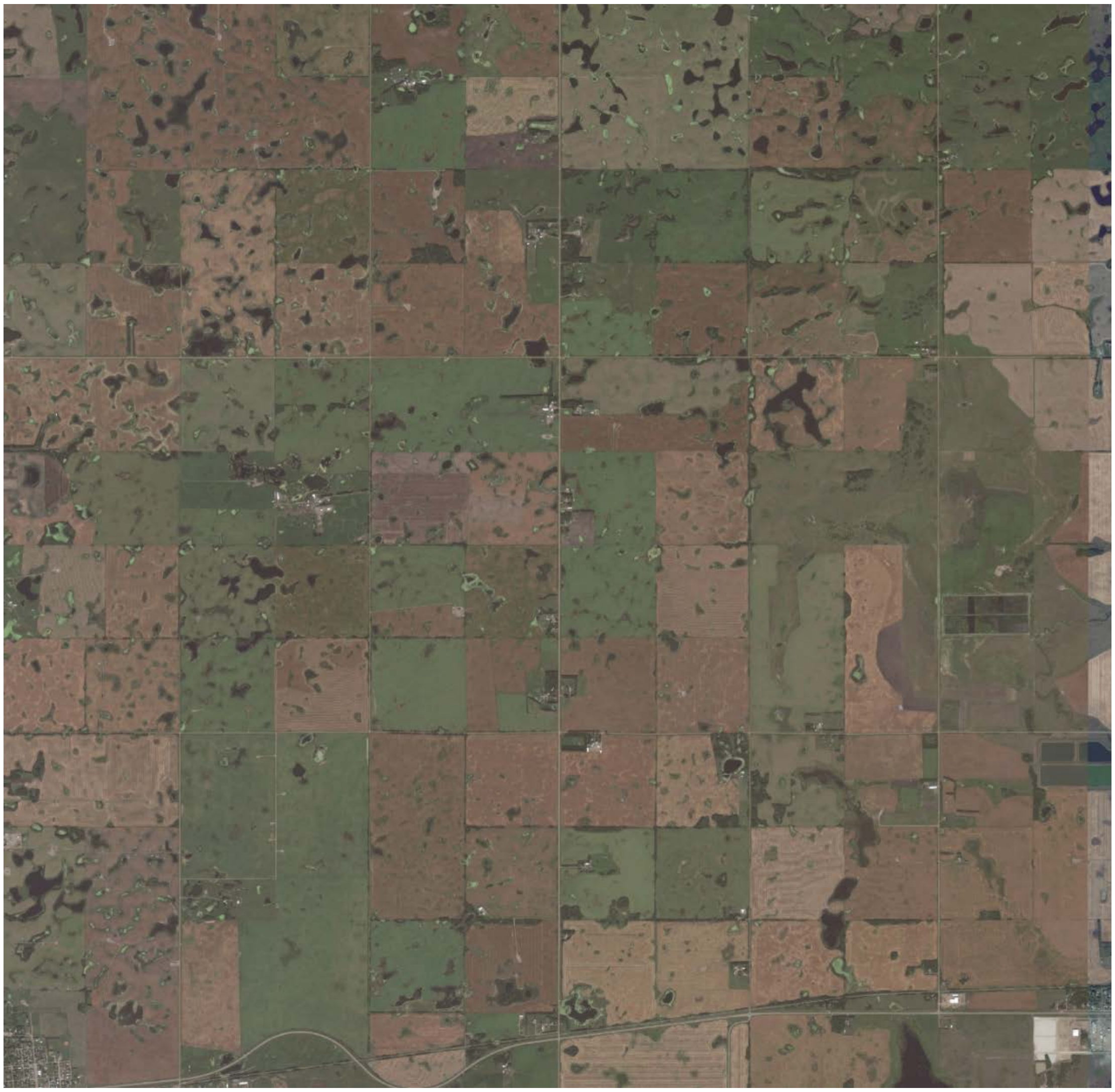

Figure 6. A LandSat photograph of land use and land cover for Parish TO39R20W4

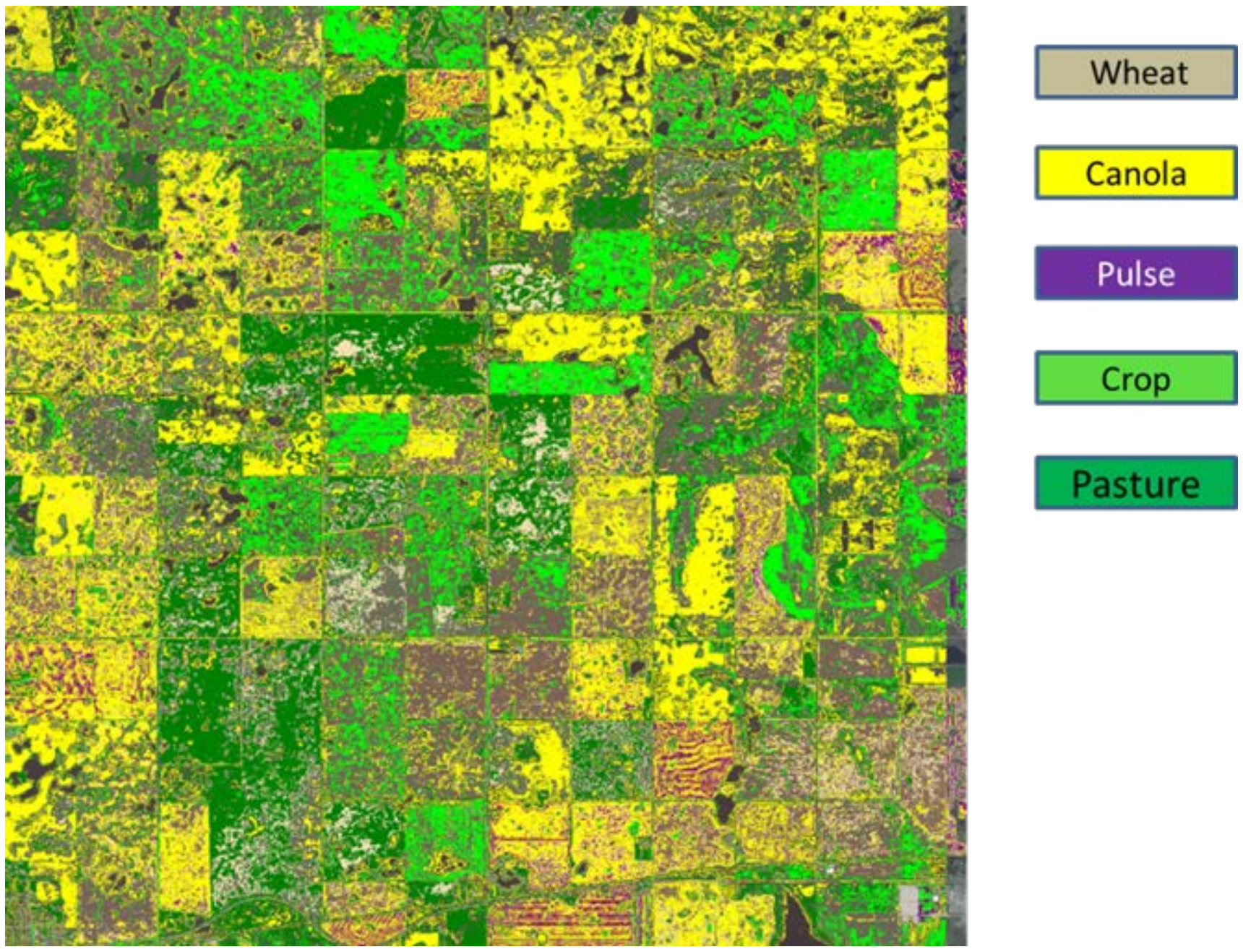

Figure 7. Maps generated by Land Analysis Plug-in by implementing Algorithm

**Table 1:** A comparison between pixel analysis data and statistical data.

| Parish/ Township | TO39R20W4 statistical Data | TO39R20W4 Analysis using Bing Aerial Image |
|---|---|---|
| Wheat | 25.69 $km^2$ | 24.03 $km^2$ |
| Canola | 23.24 $km^2$ | 21.05 $km^2$ |
| Pulses | 2.13 $km^2$ | 1.85 $km^2$ |
| Crop | 72.36 $km^2$ | 56.19 $km^2$ |
| Pasture | 12.58 $km^2$ | 14.71 $km^2$ |
| Agri Land | 84.94 $km^2$ | 70.90 $km^2$ |
| Water | NA | 1.87 $km^2$ |
| Total Area | 95.44 $km^2$ | 93.27 $km^2$ |